\documentstyle[12pt,epsfig]{article}

\newcommand{\be}{\begin{equation}}
\newcommand{\ee}{\end{equation}}
\def\aprle{\buildrel < \over {_{\sim}}}
\def\aprge{\buildrel > \over {_{\sim}}}
\begin{document}  
\topmargin 0pt
\oddsidemargin=-0.4truecm
\evensidemargin=-0.4truecm
\renewcommand{\thefootnote}{\fnsymbol{footnote}}
\newpage
\setcounter{page}{0}
\begin{titlepage}   
\vspace*{-2.0cm}  
\begin{flushright}
FISIST/6-2000/CFIF \\
hep-ph/0005173
\end{flushright}
\vspace*{0.1cm}
\begin{center}
{\Large \bf
SNO and the neutrino magnetic moment solution \\
\vspace{0.1cm}
of the solar neutrino problem} \\
\vspace{0.6cm}

{\large 
E. Kh. Akhmedov\footnote{On leave from National Research Centre Kurchatov 
Institute, Moscow 123182, Russia. E-mail: akhmedov@cfif.ist.utl.pt} and
Jo\~{a}o Pulido\footnote{E-mail: pulido@beta.ist.utl.pt}}\\  
\vspace{0.05cm}
{\em Centro de F\'\i sica das Interac\c c\~oes Fundamentais (CFIF)} \\
{\em Departamento de F\'\i sica, Instituto Superior T\'ecnico }\\
{\em Av. Rovisco Pais, P-1049-001 Lisboa, Portugal}\\  
\end{center}
\vglue 0.6truecm
\begin{abstract}
Assuming that the solar neutrino deficit observed in the Homestake,
SAGE, Gallex, Kamiokande and Super-Kamiokande experiments is due to the 
interaction of the neutrino transition magnetic moment with the solar
magnetic field, we calculate the expected values of a number of observables 
to be measured by SNO. For three model solar magnetic field profiles that
produce the best fits of the previous data we calculate the charged current
event rate, the (neutral current)/(charged current) event ratio and the
charged current electron spectrum as well as its first and second moments.
We study the dependence of the calculated observables on the choice of
the magnetic field profile and on the value of the solar $hep$ neutrino   
flux. We also compare our results with those obtained assuming that the solar 
neutrino problem is due to neutrino oscillations in vacuum or in matter. We 
show that there is an overlap or partial overlap between our predictions
and those found for each of the oscillation solutions (SMA, LMA, LOW and VO).
Given the uncertainties in the calculations and the expected uncertainties
in the experimental results, the unambiguous discrimination between
the two types of solutions to the solar neutrino problem (neutrino    
oscillations and magnetic moments) on the basis of the average rates and
electron spectrum distortions appears to be difficult. The possible time
dependence of the charged current signal and spectrum distortion in
the case of the magnetic moment solution therefore remains the best hope
for such a discrimination. For a hybrid solution (neutrino magnetic moment
plus flavour mixing) the smoking gun signature would be an observation of
$\bar{\nu}_e$'s from the sun.
\end{abstract}
\end{titlepage}   
\renewcommand{\thefootnote}{\arabic{footnote}}
\setcounter{footnote}{0}
\section{Introduction}

The Sudbury Neutrino Observatory (SNO) \cite{SNO} 
is capable of detecting solar neutrinos through charged current and neutral 
current neutrino deuteron reactions as well as through $\nu e$ scattering. 
It is expected to perform precision measurements of a number of the 
characteristics of the solar neutrino flux.  The forthcoming data from SNO 
will complement in a very important way the already available data of 
Homestake, SAGE, GALLEX, Kamiokande and Super-Kamiokande experiments 
\cite{previous}, allowing crucial tests of the proposed solutions of the 
solar neutrino problem.  

Recently, a comprehensive new study of the possible implications of the 
forthcoming SNO data for neutrino oscillation solutions of the solar neutrino 
problem has been performed by Bahcall, Krastev and Smirnov (hereafter BKS) 
\cite{BKS3,BKS4}. In the present paper we analyse possible implications
of the SNO measurements for another particle physics solution of the solar 
neutrino problem which is currently consistent with all the data -- the 
neutrino magnetic moment scenario. We also compare our predictions with those 
of BKS and discuss the possibilities of discriminating between the two 
scenarios. 

Neutrinos with transition magnetic moments experience a simultaneous rotation 
of their spin and flavour in external magnetic fields (spin-flavour
precession) \cite{SV,VVO}. This precession can be resonantly enhanced 
in matter \cite{LM,Akh1}. The resonance spin-flavour precession of 
neutrinos (RSFP) in the matter and magnetic field of the sun can efficiently 
transform solar $\nu_{eL}$ into, e.g., $\nu_{\mu R}$ (or $\bar\nu_{\mu R}$ 
in the case of the transition magnetic moment of Majorana neutrinos) 
provided that the neutrino magnetic moment $\mu_\nu\aprge 10^{-11} \mu_B$ 
and the average strength of the solar magnetic field in the region where 
the transition takes place $B \aprge 40$ kG. As $\nu_{\mu R}$ or 
$\bar{\nu}_{\mu R}$ either do not interact with the detectors or interact
with them more weakly than $\nu_{eL}$, RSFP can account for the deficiency
of observed solar neutrino flux. Analyses performed in the framework of this 
scenario \cite{ALP1,Pul,LN,GN,PA} show that a very good fit of the
currently available data can be achieved. In particular, the fits of the 
total detection rates are much better than those in the case of the  
Mikheyev-Smirnov-Wolfenstein (MSW \cite{MSW}) effect whereas the fits of 
the recoil electron spectrum in Super-Kamiokande in the frameworks of the 
RSFP and MSW effect are of nearly the same quality \cite{GN,PA}. 

In the present paper we calculate, within the RSFP mechanism, the expected
values of a number of observables to be measured by SNO. For three model 
solar magnetic field profiles that produce the best fits of the previous 
data we calculate the charged current event rate, the 
(neutral current)/(charged current) event ratio and the
charged current electron spectrum as well as its first and second moments.
We study the dependence of the calculated observables on the choice of
the magnetic field profile and on the value of the solar $hep$ neutrino   
flux. We then compare our results with those obtained by BKS assuming
that the solar neutrino problem is due to  neutrino oscillations in vacuum 
or in matter. 

The implications of the RSFP of solar neutrinos for the SNO experiment have 
been analysed in the past \cite{past,BG1}. Our analysis is more detailed,
especially as far as the charged current electron spectrum is concerned; 
in addition, we use the magnetic field profiles which provide a good fit of 
all existing solar neutrino data, including the recent Super-Kamiokande ones 
which were not available when the analyses of \cite{past,BG1} were performed. 

The paper is organized as follows: 
in sec. 2 we present the three magnetic field profiles to be used, in 
sec. 3 we define and evaluate the physical quantities which are relevant for 
the SNO experiment, for both the charged current (CC) and neutral current 
(NC) $\nu d$ reactions. 
Finally in section 4 we discuss the obtained results and draw our main
conclusions.

\section{Solar magnetic field profiles}

Unfortunately, very little is known about the inner magnetic field of the 
sun, and one is forced to use various model magnetic filed profiles in 
order to account for the solar neutrino data in the framework of the RSFP 
mechanism. In a previous paper \cite{PA} we investigated seven different 
magnetic field profiles and found that only some of them produce acceptable 
fits of the data. Here we present the three profiles which produced the best 
fits and which will be used in the present study. They are shown in figs. 
1 and 2 of ref. \cite{PA} (notice that our present profiles 1, 2 and 3 are
respectively profiles 2, 5 and 6 of ref. \cite{PA}).  

All three profiles show a sudden rise around the bottom of the convective 
zone, at 0.65-0.71 of the solar radius followed by a smoother decrease up to 
the surface. Here the field intensity is at most of the order of a few hundred 
Gauss. The best fits quoted here were taken from 
ref. \cite{PA} and correspond to a neutrino magnetic moment $\mu_{\nu} = 
10^{-11}\mu_B$ \footnote{We recall that only the product of the magnetic 
moment and the magnetic field enters in the neutrino evolution equation
and so our results apply to any other value of ${\mu_{\nu}}$ provided
that the magnetic field is rescaled accordingly.}. 
The first profile (profile 1) is \cite{ALP1} 
\be
B=0\,,~~~~~~~x<x_R.
\ee
\be
B=B_0\frac{x-x_R}{x_C-x_R}\,,~x_{R}\leq x\leq x_{C}
\ee
\be
B=B_0\left[1-\frac{x-x_C}{1-x_C}\right]\,,~x_{C}<x\leq 1.
\ee
Here $x=r/R_\odot$, $R_\odot$ being the solar radius, and we take 
$x_R=0.65$, $x_C=0.80$. The best fit for the Homestake, SAGE, GALLEX, 
Kamiokande and Super-Kamiokande rates in this case corresponds to the values 
of the mass squared difference and peak field strength, respectively, 
$\Delta m^2_{21}=1.20\times10^{-8}$ eV$^2$ and 
$B_0=1.23 \times10^5$ G ($\chi^2/d.o.f.=0.10/1$).

The next profile (profile 2) \cite{AB,ALP1} is 
\be
B=0\,,~~~~~~x<x_{R}
\ee
\be
B=\frac{B_0}{\cosh30(x-x_R)}\,,~x\geq x_{R}
\ee
with the best fit at $\Delta m^2_{21}=2.1\times10^{-8}$ eV$^2$, $B_0=1.45 
\times10^5$ G ($\chi^2/d.o.f.=0.055/1$). 

Profile 3 \cite{Tom} is a modification of the previous one: 
\be
B=2.16\times10^3~{\rm G}\,,~~~~x\leq 0.7105
\ee
\be
B=B_{1}\left[1-\left(\frac{x-0.75}{0.04}\right)^2\right]\,,~0.7105<x<0.7483
\ee
\be
B=\frac{B_{0}}{\cosh30(x-0.7483)}\,,~0.7483\leq x\leq 1
\ee
with $B_0=0.998B_1$. The best fit is in this case is $\Delta m^2_{21}=1.6 
\times10^{-8}$ eV$^2$, $B_0=9.6\times10^4$ G with $\chi^2/d.o.f.=0.048/1$ 
\cite{PA}. The best fits of the Super-Kamiokande recoil electron spectrum 
found for these three profiles, at values of $\Delta m^2_{21}$ and 
magnetic field strength rather close to those indicated above, were 
respectively $\chi^2=23.9, 23.5, 23.6$ for 16 d.o.f. \cite{PA}. In what 
follows we will be using the rate best fits. 

We have determined the areas on the plane $\Delta m^2_{21}$, $B_0$ 
corresponding to the 90\% c.l. fits of the data ($\chi^2=\chi^2_{min}+2.7$ 
for 1 d.o.f.) for each of the three magnetic field profiles described above. 
The resulting areas are shown in fig. 1 together with the best fit values.     

\begin{figure}
\setlength{\unitlength}{1cm}
\begin{center}
\vspace{-1.0cm}
\epsfig{file=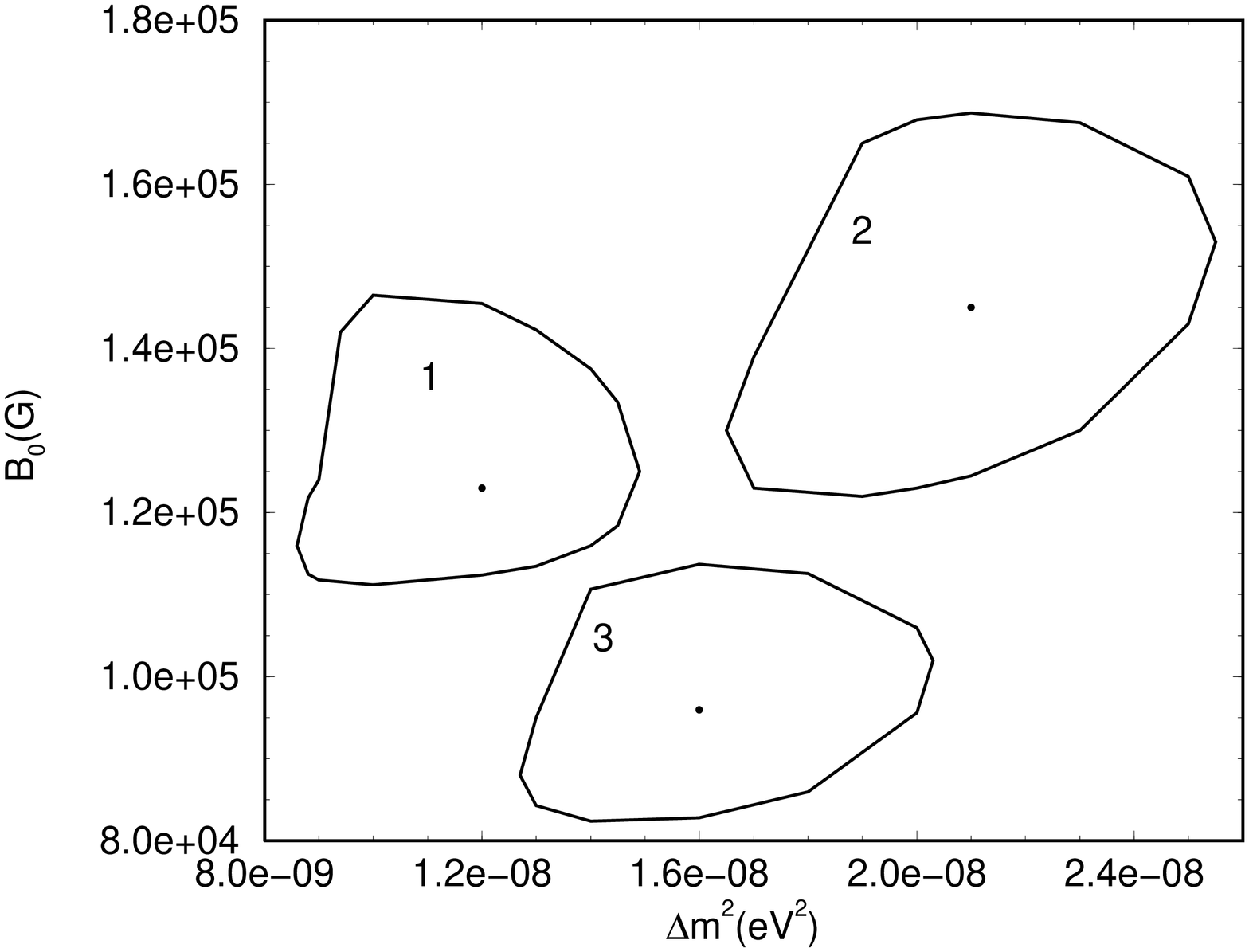,width=6.0cm,angle=270}
\end{center}
\vspace{-0.4cm}
%
{\small Figure 1:  
Regions in the $(\Delta m_{21}^2,\,B_0)$ plane allowed at 90\% c.l. by the 
combined fit of the Homestake, SAGE, Gallex, Kamiokande and Super-Kamiokande 
detection rates for the three magnetic field profiles used in the present 
study. Dots correspond to best fits.}
\end{figure}

\section{Charged and neutral currents at SNO}

SNO can detect solar neutrinos through the CC reaction  
\be
\nu_e+d \rightarrow p+p+e^- 
\label{CC}
\ee
(energy threshold $Q=1.44$ MeV), NC reaction 
\be
\nu_x+d \rightarrow n+p+\nu_x\, 
\label{NC}
\ee
(energy threshold equal to the deuteron binding energy $E_B=2.225$
MeV, $x=e\,,~\mu$ or $\tau$), 
and through the neutrino electron scattering $\nu_x + e \to \nu_x + e$.  
The rate and the recoil electron energy 
spectrum in the latter process at SNO are expected to be very similar to
those at Super-Kamiokande studied in \cite{PA}; for this reason we shall 
not discuss them here. 

We first consider the relative electron spectrum of reaction (\ref{CC}) 
which is defined as the ratio of the electron energy distribution and the 
corresponding distribution for standard model neutrinos. By
standard model neutrinos we mean neutrinos described by the standard
electroweak model (i.e. with no magnetic moment or mass) with their fluxes
given by the standard solar model. 
We subdivide the electron energy spectrum into 0.5 MeV bins, with the 
relative spectrum for the i$th$ bin given by 
\be
S_i=\frac{
\int_{T_i}^{T_{i+1}}dT \int_{Q}^{\infty}dE\,f(E)P(E)\int_{0}
^{E-Q}dT^{'}\frac{d\sigma_{CC}}{dT^{'}}(E,T^{'})R(T,T^{'})}
{\int_{T_i}^{T_{i+1}}dT \int_{Q}^{\infty}
dE f(E)\int_{0}^{E-Q}dT^{'}
\frac{d\sigma_{CC}}{dT^{'}}(E,T^{'})R(T,T^{'})}
\equiv\frac{\int_{T_i}^{T_{i+1}}dT \,dR_{CC}/dT}{\int_{T_i}^{T_{i+1}}dT  
\, dR^{st}_{CC}/dT}
\label{S}
\ee
Here $P(E)$ denotes the electron neutrino survival probability, $T$ is the 
measured recoil electron kinetic energy, in contrast to the physical one,
$T^{'}$. Following \cite{BL,BKS4}, we approximate the energy resolution 
function of the detector $R(T',T)$ by a Gaussian 
\be
R(T,T{'})=\frac{1}{\Delta_{T^{'}}\sqrt{2\pi}}\exp\left[-\frac{(T^{'}-T)^2} 
{2\Delta^2_{T^{'}}}\right]\,,\quad\quad 
\Delta_{T^{'}}=(1.1~{\rm MeV})\sqrt{\frac{T^{'}}{10~{\rm MeV}}}\,. 
\ee
The cross sections of the CC and NC $\nu d$ reactions used throughout this 
paper are those of Kubodera {\it et al.} \cite{KN}. 
The SNO experiment is only sensitive to the high-energy $^8$B and $hep$
components of the solar neutrino flux. We use the BP98 solar neutrino 
spectrum $f(E)$ \cite{BP98,hom} except that the $hep$ neutrino flux is 
allowed to deviate from its nominal BP98 value.  
The $hep$ flux is about 3 orders of magnitude smaller than that of $^8$B 
neutrinos and so gives a negligible contribution to the CC and NC rates.
However, its contribution to the highest energy part of the CC electron 
spectrum may be important as its endpoint energy is 18.8 MeV whereas that of 
$^8$B neutrinos is only 15 MeV.   

Unless otherwise specified, we consider the case of Majorana neutrinos 
in the present paper. The survival probability $P(E)$ is calculated by 
solving the RSFP evolution equation numerically \cite{rev}. 
The relative electron spectra for the three magnetic field 
profiles discussed in sec. 2 are shown in fig 2. The highest energy bin
($E_e\ge 14$ MeV) represents the average value of $S$ for measured electron 
energies between 14 and 20 MeV. The solid lines correspond to the best
fits whereas the dashed and dotted lines show the maximum and minimum
values obtained within the 90\% c.l. allowed regions of values of $\Delta
m^2_{21}$ and $B_0$ obtained by fitting the previous data and shown in
fig. 1. In order to check the sensitivity of the spectrum to the poorly
known $hep$ flux we perform all the calculations for three values of
the flux scaling factor, $f_{hep}=0$, 1 and 20, with $f_{hep}=1$
corresponding to the nominal $hep$ flux value in the BP98 standard solar 
model (SSM).

We also calculate the first and second moments $\langle T\rangle$ and 
$\sigma$ of the electron spectrum, defined through 
\be
\langle T^n\rangle=\frac{1}{R_{CC}}\int_{T_{m}}^{\infty} T^n \,
\frac{dR_{CC}}{dT} \,dT\,, \quad \quad
R_{CC}=\int_{T_m}^\infty \frac{dR_{CC}}{dT}\, dT\,,\quad\quad 
\sigma =\sqrt{\langle T^2\rangle -\langle T\rangle^2}\,,
\label{T}
\ee
with 
$dR_{CC}/dT$ defined in (\ref{S}). The deviations of these moments from 
their values predicted by the SSM are convenient quantitative 
characteristics of the electron spectrum distortion \cite{BKL}. 
The results of the calculations for two different values of 
the threshold electron kinetic energies $T_m$ which correspond to the 
total energies of 5 MeV and 8 MeV and for $f_{hep}=1$ and 20 are presented 
in tables I and II. We have also calculated $\langle T\rangle$ and $\sigma$
for $f_{hep}=0$ and found that their values are practically indistinguishable 
from those for $f_{hep}=1$. This also applies to the CC event ratio 
$r_{CC}$ and NC/CC double ratio $\bar{r}_{NC}$ discussed below. 
The errors presented in tables I and II include 
the 90\% c.l. errors coming from the fits of the parameters $\Delta
m^2_{21}$ and $B_0$ (fig. 1) as well as from the uncertainties in the 
energy resolution and scale, $^8$B neutrino spectrum, reaction cross
section and statistics (assuming 5000 CC events). These uncertainties were 
taken from table II of ref. \cite{BKS4}.  

\begin{table}  
{\small Table I: First moment $\langle T\rangle$ of the electron spectrum 
in the CC reaction (\ref{CC}) (in MeV) for three different magnetic field
profiles and for standard model neutrinos. The results are presented for 
two values of the electron threshold energy $E_e=5$ MeV and $E_e=8$ MeV 
and two values of the $hep$ neutrino flux scaling factor, $f_{hep}=1$ and 
$f_{hep}=20$ (indicated in the parentheses in the first line). The errors 
correspond to 90\% c.l. (see the text for details). }
\begin{center}
\begin{tabular}{ccccc} \hline \hline
Profile & $\langle T\rangle$~(5 MeV; 1) & 
$\langle T\rangle$~(8 MeV; 1) & $\langle T\rangle$~(5 MeV; 20) & $\langle
T\rangle$~(8 MeV; 20) \\ \hline 
1 & $7.477\pm\,^{0.124}_{0.126}$ & $9.117\pm 0.150$ &
$7.541\pm\,^{0.126}_{0.127}$ & $9.198 \pm \,^{0.151}_{0.152}$ \\ 
2   & $7.491\pm 0.125$ & $9.121 \pm 0.150$ & $7.556 \pm 0.126$ & $9.202
\pm 0.152$ \\  
3 & $7.484\pm 0.125$ & $9.119 \pm 0.150$ & $7.549 \pm 0.126$ & $9.200 \pm
0.152$ \\ 
Stand. $\nu$'s & $7.423\pm 0.122$ & $9.101\pm 0.150$ & $7.484 \pm
0.123$ & 
$9.179 \pm 0.151$ \\ \hline
\end{tabular}
\end{center}
\end{table}

\begin{table}  
{\small Table II: Same as in table I but for the second moment of 
the electron spectrum $\sigma$.}
\begin{center}
\begin{tabular}{ccccc} \hline \hline
Profile & $\sigma$~(5 MeV; 1) & 
$\sigma$~(8 MeV; 1) & $\sigma$~(5 MeV; 20) & $\sigma $~(8 MeV; 20) \\
\hline 
1 & $1.880\pm 0.070$ & $1.273 \pm 0.047$ &
$1.953\pm\,^{0.072}_{0.073}$ & $1.385 \pm \,^{0.051}_{0.052}$ \\ 
2   & $1.881\pm 0.070$ & $1.274 \pm 0.047$ & $1.955 \pm 0.072$ & $1.387 
\pm 0.051$ \\  
3 & $1.881\pm 0.070$ & $1.273 \pm 0.047$ & $1.954 \pm 0.072$ & $1.386 \pm
0.051$ \\ 
Stand. $\nu$'s & $1.874\pm 0.069$ & $1.267\pm 0.047$ & $1.943 \pm
0.072$ & $1.375 \pm 0.051$ \\ \hline
\end{tabular}
\end{center}
\end{table}

Next we examine the ratio of the CC event rate for neutrinos undergoing RSFP 
and that for standard model neutrinos,  
\be
r_{CC}=\frac{\int_{T_m}^{\infty}(dR_{CC}/dT)\, dT}
{\int_{T_m}^{\infty}(dR^{st}_{CC}/dT)\, dT}=\frac{R_{CC}}{R_{CC}^{st}}\,.
\label{rCC}
\ee
This ratio depends on the chosen threshold energy $T_{m}$ and we again use 
the same values of $T_m$ that we used in calculating the moments of the 
electron spectrum. The results are presented in table III. The absolute 
values of the CC event rates for standard model neutrinos for the total 
electron threshold energies $E_e=5$ MeV are 5.132 SNU and 5.253 SNU for 
$f_{hep}=1$ and 20 respectively, and those for the electron threshold energy 
$E_e=8$ MeV are 2.355 SNU and 2.451 SNU. 

\begin{table} 
{\small Table III: Ratio $r_{CC}$ of the event rate due to the CC reaction 
(\ref{CC}) for neutrinos undergoing RSFP and that for standard model 
neutrinos. The values of the electron threshold energy and 
of the $hep$ neutrino flux scaling factor $f_{hep}$ 
are indicated in the parentheses in the first line. 
The errors correspond to 90\% c.l. (see the text for details). The errors 
in the parentheses do not include the $^8$B neutrino flux uncertainties. }
\begin{center}
\begin{tabular}{ccccc} \hline \hline
Profile & $r_{CC}$~(5 MeV; 1) & 
$r_{CC}$~(8 MeV; 1) & $r_{CC}$~(5 MeV; 20) & $r_{CC}$~(8 MeV; 20) \\
\hline 
1 & $0.421\pm^{0.165(0.108)}_{0.121(0.050)}$ & \vspace*{0.2cm} 
$0.431 \pm ^{0.169(0.111)}_{0.124(0.051)}$ & 
$0.432\pm^{0.166(0.105)}_{0.124(0.049)}$ & 
$0.450\pm^{0.177(0.110)}_{0.143(0.051)}$ \\ 
2 & $0.406\pm^{0.136(0.064)}_{0.116(0.046)}$ & \vspace*{0.2cm} 
$0.415\pm^{0.139(0.066)}_{0.119(0.047)}$ & 
$0.419\pm^{0.139(0.063)}_{0.119(0.046)}$ & 
$0.436\pm^{0.145(0.066)}_{0.124(0.049)}$ \\  
3 & $0.415\pm^{0.139(0.064)}_{0.119(0.048)}$ & 
$0.426\pm^{0.142(0.066)}_{0.123(0.050)}$ & 
$0.428\pm^{0.142(0.064)}_{0.122(0.048)}$ & 
$0.446\pm^{0.148(0.067)}_{0.128(0.050)}$  \\
\hline
\end{tabular}
\end{center}
\end{table}

We have also calculated the total event rate of the NC reaction (\ref{NC}) 
\be
R_{NC}=\int_{E_B}^{\infty}f(E)\{P(E)\,\sigma_{NC}^{\nu d}(E)+
[1-P(E)]\,\sigma_{NC}^{\bar{\nu} d}(E)\}\epsilon(E)\,dE
\label{RNC}
\ee
Here $\epsilon(E)$ is the NC detection efficiency; following refs. 
\cite{BL,BKS4} we have taken $\epsilon(E)=0.50$. For
neutrino energies $E\le 15$ MeV the cross sections of the NC $\nu d$ 
and $\bar{\nu} d$ reactions differ by less than 7.3\%, the difference 
at $E=9$ MeV being about 4\% \cite{KN}. These differences are 
within the uncertainty of the value of the NC cross sections itself, and
to a good approximation one can put $\sigma_{NC}^{\bar{\nu}d}(E)=
\sigma_{NC}^{\nu d}(E)$. The NC rate (\ref{RNC}) then does not depend on the 
$\nu_e$ survival neutrino probability $P(E)$, and one therefore expects
$R_{NC}$ in the case of neutrinos undergoing RSFP to coincide with that 
for standard neutrinos $R_{NC}^{st}$ \cite{Akh2,BG1}. Notice that for 
$\mu_\nu\aprle 10^{-11}\mu_B$ the electromagnetic contribution to the NC 
neutrino-deuteron disintegration reaction (\ref{NC}) due to the neutrino 
magnetic moment is more than eight orders of magnitude smaller than the
standard electroweak one \cite{ABer} and so can be safely neglected. For 
$f_{hep}=1$ and 20 we obtain $R_{NC}^{st}=1.216$ SNU and 1.242 SNU
respectively. 

\begin{table} 
{\small Table IV: Double ratio $\bar{r}_{NC}=r_{NC}/r_{CC}$.  
The values of the electron threshold energy for the CC reaction and 
of the $hep$ neutrino flux scaling factor $f_{hep}$ 
are indicated in the parentheses in the first line. 
The errors correspond to 90\% c.l. (see the text for details). }
\begin{center}
\begin{tabular}{ccccc} \hline \hline
Profile & $\bar{r}_{NC}$~(5 MeV; 1) & 
$\bar{r}_{NC}$~(8 MeV; 1) & $\bar{r}_{NC}$~(5 MeV; 20) & $\bar{r}_{NC}$~(8
MeV; 20) \\
\hline 
1 & $2.375\pm^{0.204}_{0.576}$ & \vspace*{0.2cm} 
$2.320 \pm ^{0.200}_{0.566}$ & 
$2.315\pm^{0.184}_{0.533}$ & 
$2.222\pm^{0.177}_{0.511}$ \\ 
2 & $2.463\pm^{0.198}_{0.331}$ & \vspace*{0.2cm} 
$2.410\pm^{0.187}_{0.329}$ & 
$2.387\pm^{0.174}_{0.303}$ & 
$2.294\pm^{0.165}_{0.296}$ \\  
3 & $2.410\pm^{0.199}_{0.320}$ & 
$2.347\pm^{0.196}_{0.308}$ & 
$2.336\pm^{0.180}_{0.296}$ & 
$2.242\pm^{0.173}_{0.284}$  \\
\hline
\end{tabular}
\end{center}
\end{table}

An important characteristics of the $\nu d$ reactions at SNO is the ratio 
$\bar{r}_{NC}$ of the ratios of the NC and CC event rates to their 
respective values for standard neutrinos. This double ratio is free of
many uncertainties which are present in the single ratios $r_{CC}$ and 
$r_{NC}\equiv R_{NC}/R_{NC}^{st}$. In particular, the errors in $r_{CC}$ 
and $r_{NC}$ due to the uncertainties in the flux of $^8$B neutrinos 
practically cancel out in the double ratio, and those due to the 
uncertainties in the CC and NC cross sections cancel out to a large
extent. As was mentioned above, for Majorana neutrinos one has $r_{NC}=1$ to 
a good accuracy, and so $\bar{r}_{NC}=1/r_{CC}$. The values of $\bar{r}_{NC}$ 
for the three magnetic field profiles that we consider are given in table IV. 
The indicated errors are the combined 90\% c.l. ones coming from the 
uncertainties in the fitted values of $\Delta m^2_{21}$ and $B_0$ (see fig. 1) 
as well as from the uncertainties in the electron energy resolution and scale 
in the CC reaction, $^8$B neutrino spectrum, CC and NC cross sections and
statistics (assuming 5000 CC events). The latter uncertainties were taken 
from table II of ref. \cite{BKS4}. 

Finally, we remark on the case of Dirac neutrino transition moments. In a
sense this case is similar to neutrino oscillations into sterile neutrinos 
because the RSFP due to Dirac neutrino magnetic moments transforms
$\nu_{eL}$ into sterile $\nu_{\mu R}$ or $\nu_{\tau R}$. The CC rate ratio  
$r_{CC}$ in this case is similar to that in the Majorana neutrino case 
since the survival probabilities in the two cases are very close to each 
other. However, the NC rate $R_{NC}$ and the rate ratio $r_{NC}$ (and so 
the double ratios $\bar{r}_{NC}$) in the Majorana and Dirac cases are
drastically different: while for Majorana neutrinos $R_{NC}\simeq R_{NC}^{st}$  
($r_{NC}\simeq 1$), in the case of Dirac neutrinos $R_{NC}$ exhibits 
a suppression similar to that of $R_{CC}$ and so the double ratio  
$\bar{r}_{NC}$ is close to unity. Thus, the double ratio $\bar{r}_{NC}$ 
can be used to discriminate between Dirac and Majorana neutrinos. 
We would like to stress that this is one of a very few known quantities 
that hold such a potential, the best known other example being the 
neutrinoless double beta decay. Notice that in general the RSFP of Dirac 
neutrinos leads to a fit of the solar neutrino data that is worse than 
that due to the RSFP of Majorana neutrinos and therefore we do not pursue 
here this possibility in detail. 

\section{Summary and discussion}
We have investigated the expectations from the RSFP solution to the solar 
neutrino problem for the SNO experiment using the solar magnetic field
profiles that provide the best fits for the rates of the Homestake, SAGE, 
GALLEX, Kamiokande and Super-Kamiokande experiments. We considered Majorana 
neutrinos and discussed the quantities that are relevant for the CC and 
NC processes (\ref{CC}) and (\ref{NC}) in SNO. To this end we have 
examined the expected relative CC electron spectrum $S(E_e)$ as well as
the first and second moments of the spectrum $\langle T\rangle$ and $\sigma$, 
the ratio $r_{CC}$ of the CC event rate to that of standard model neutrinos, 
and the ratio $\bar{r}_{NC}$ of the ratios of the NC and CC event rates 
to their respective values for standard neutrinos. 

The values of the CC event rate ratio $r_{CC}$ calculated for the electron
threshold energies 5 MeV and 8 MeV are very close to each other, the 
difference being only 2 -- 4\% (table III). This is related to the the
fact that, for the magnetic field profiles that we consider, the high-energy 
part of the survival probability $P(E)$ for neutrinos undergoing the RSFP is 
rather flat (see figs. 3 and 4 in \cite{PA}). Therefore both $R_{CC}$ and 
$R_{CC}^{st}$ decrease with increasing electron energy threshold to nearly 
the same extent so that their ratio $r_{CC}$ changes very little. For the
same reason the double ratio $\bar{r}_{NC}$, which depends on the CC electron 
threshold energy only through $r_{CC}$, is rather insensitive to the value of 
this threshold energy. In contrast to this, the first and second moments of 
the CC electron spectrum $\langle T\rangle$ and $\sigma$ depend sensitively on 
the chosen electron threshold energy (see tables I and II). 

To assess the sensitivity of the observables to be measured by SNO to the 
poorly known value of the $hep$ neutrino flux we have performed all the 
calculations for three values of this flux: zero flux ($f_{hep}=0$), the
nominal flux of the BP98 standard solar model ($f_{hep}=1$) and a factor
of 20 larger one ($f_{hep}=20$). Changing the value of $f_{hep}$ from 0 
to 1 does not lead to any noticeable difference in the calculated
observables. The CC event rate ratio $r_{CC}$ and the NC/CC double ratio
$\bar{r}_{CC}$ only weakly depend on the value of the $hep$ flux. This is 
mainly because the expected contributions of the $hep$ neutrinos to the CC 
and NC event rates $R_{CC}$ and $R_{NC}$ are very small and even increasing 
these contributions by a factor of 20 would not change $R_{CC}$ and $R_{NC}$ 
much. The sensitivity to $f_{hep}$ of the ratio $r_{CC}$ and double ratio 
$\bar{r}_{NC}$ is further reduced due to a partial cancellation of the 
$f_{hep}$ dependences of the numerators and denominators. 
The first moment of the CC electron spectra $\langle T\rangle$ is also
rather insensitive to the $hep$ neutrino flux: 
changing $f_{hep}$ from 1 to 20 modifies  $\langle T\rangle$ by less than 
1\%. The second moment $\sigma$ is more sensitive to the $hep$ neutrino flux: 
changing $f_{hep}$ from 1 to 20 increases $\sigma$ by about 4\% for the 
electron energy threshold $E_{min}=5$ MeV and by about 9\% for $E_{min}=8$ MeV. 
These features are independent of whether or not neutrinos undergo 
RSFP. 

\begin{figure}
\setlength{\unitlength}{1cm}
\begin{center}
\vspace{-0.99cm}
\hspace*{-1.8cm}
\epsfig{file=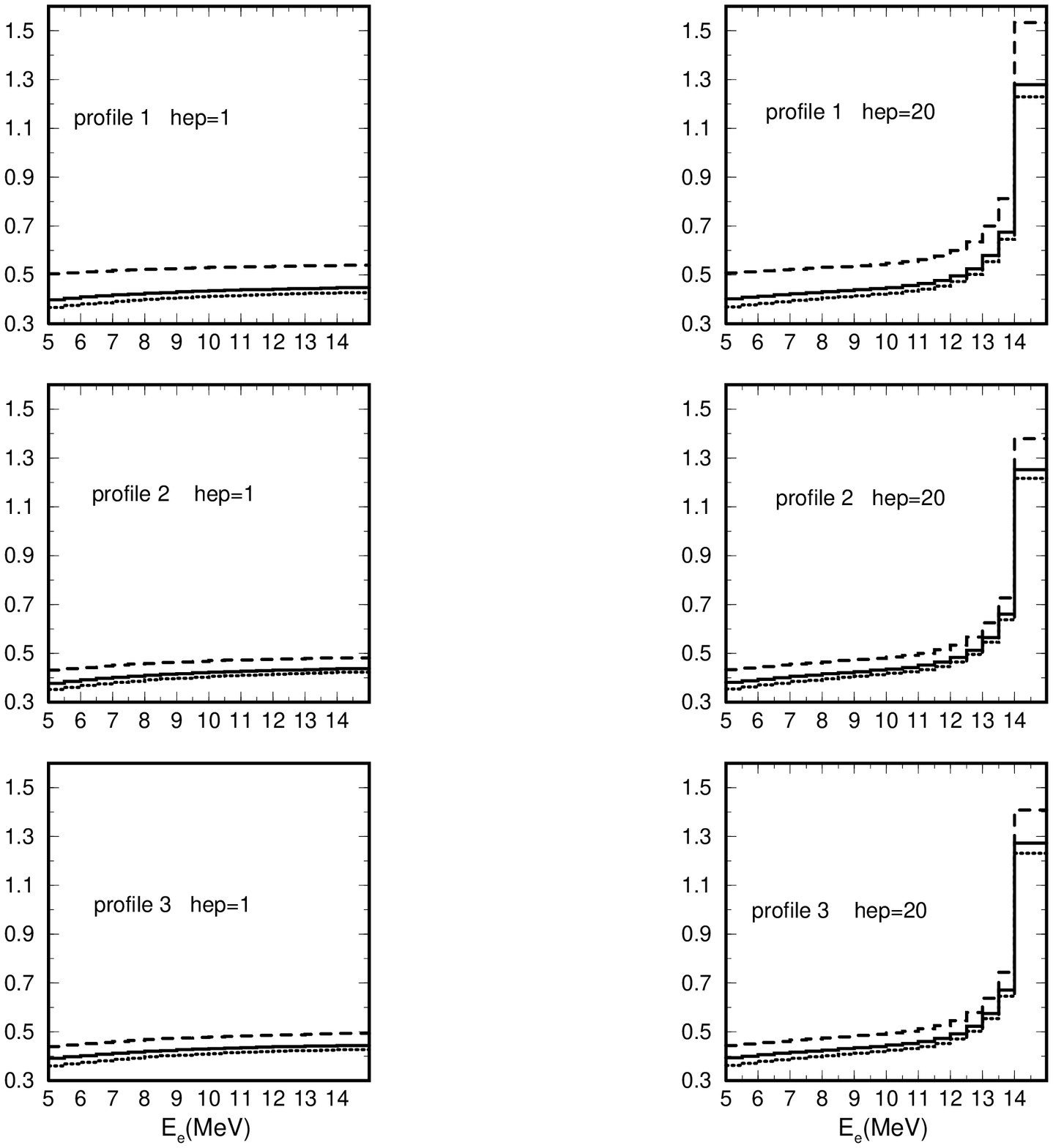,height=12.0cm,angle=270}
\end{center}
\vspace{-0.2cm}
{\small Figure 2: Relative CC electron spectra $S(E_e)$. 
Solid lines -- best fits, dashed and dotted lines -- maximum and minimum
values corresponding to the allowed regions of fig. 1.} 
\end{figure}
As can be seen from fig. 2, the high energy part of the relative CC
electron spectrum ($E\aprge 12$ MeV) depends sensitively on the
magnitude of the $hep$ neutrino flux. 
For $f_{hep}=20$ the excess of the number of the high energy events 
($E\ge 14$ MeV) over the SSM prediction with nominal $hep$ flux can be
quite significant. It should be noted, however, that, although it is quite
likely that the actual $hep$ neutrino flux exceeds the nominal one of the
BP98 model, the illustrative value $f_{hep}=20$ that we used 
may in fact be too high. The most recent calculation \cite{hep}
gives $f_{hep}\approx 5$.  

All quantities that we have calculated have very similar values for the 
three magnetic field profiles that we chose. 
In particular, the values $\langle T\rangle$ and $\sigma$ differ by less
than 1\% for different profiles, and those of $r_{CC}$ and $\bar{r}_{NC}$ 
differ by 3 - 4\%.  This is the consequence of the fact that the magnetic 
field profiles used in our calculations, although rather different, lead to 
very similar $\nu_e$ survival probabilities $P(E)$ \cite{PA}. 

We shall now compare our predictions with those for standard neutrinos 
(with no magnetic moment or mass) and for neutrinos undergoing oscillations 
in vacuum or in matter. As can be seen from table III, the CC detection rates 
for neutrinos undergoing RSFP constitute about 40\% of those for the standard 
neutrinos and SSM fluxes. The errors in $r_{CC}$ are rather large (90\%
c.l. errors $\sim 35 - 40\%$), with about a half of their values coming from 
the uncertainty of the $^8$B neutrino flux. The predicted values of $r_{CC}$ 
differ from the SSM one $r_{CC}=1$ by about $5\sigma$. However, 
for neutrinos of vanishing mass and magnetic moments, the $^8$B
neutrino flux measured by Super-Kamiokande, $\phi_{^8{\rm B}}=(0.475\pm
0.015) \phi_{^8{\rm B}}^{\rm SSM}$, implies that the actual $^8$B neutrino
flux is about 0.48 of the one given by the SSM and so the expected value
of the CC event rate ratio for standard  
neutrinos is $r_{CC}\simeq 0.48$ rather than $r_{CC}=1$. This value is 
less than $1\sigma$ away from our predictions given in table III, and so 
$r_{CC}$ is not a suitable parameter for discriminating between standard
and non-standard neutrinos. 

The NC/CC double ratio $\bar{r}_{NC}$ has smaller errors (except for 
profile 1); it is practically independent of the uncertainty of the $^8$B 
neutrino flux and in addition a number of other uncertainties of $R_{CC}$ and 
$R_{NC}$ drop out from this ratio. The calculated values of $\bar{r}_{NC}$ are 
of the order 2.5 and exceed the prediction for standard neutrinos 
$\bar{r}_{NC}=1$ by more than $7\sigma$ for profiles 2 and 3 and by about 
$4\sigma$ for profile 1. Thus $\bar{r}_{NC}$ is an ideal indicator of 
non-standard neutrinos. 

The predicted relative CC electron spectrum $S(E_e)$ for $f_{hep}=1$ is
rather flat (see fig. 2). For $f_{hep}=20$, one expects an excess of the 
high energy events. In both cases there is also a small decrease at low
energies, $E\aprle 8$ MeV. The shapes of the relative spectrum that we
obtained for all three studied magnetic field profiles are very similar to
each other and to that predicted for the SMA solution of the solar neutrino 
problem in the case of neutrino oscillations (see fig. 3 in \cite{BKS4}). 
They are also similar to the $S(E_e)$ shape in the case of VAC$_{\rm S}$
oscillation solution, but the latter has a steeper increase at $E_e\aprge 
10$ MeV. The shapes of the relative CC electron spectrum for LOW and 
LMA and VAC$_{\rm L}$ solutions differ from ours in that the former two 
are almost horizontal at low energies, whereas the latter one has a
distinct dip in the energy range $E_e\simeq 8$ -- 10 MeV.  

Our predicted values of $r_{CC}$ are typically slightly larger than those 
obtained in \cite{BKS4} for the LMA and LOW solution of the solar neutrino
problem in the neutrino oscillation scenario, and in most part of the
allowed range, also larger than those of the VAC$_{\rm S}$ solution
studied by BKS (compare our table III with table V of \cite{BKS4}). Our
values of $r_{CC}$ are rather close to those for the SMA and VAC$_{\rm L}$ 
neutrino oscillation solutions. However, even in the case of the LMA and LOW  
solutions, there is a partial overlap with the RSFP predictions because the 
allowed regions of $r_{CC}$ are rather large. 
The values of the NC/CC double ratio $\bar{r}_{NC}$ that we obtained are
typically larger than those for the VAC$_{\rm L}$ solution, lower than those 
for the LMA and LOW solutions and similar to those for the SMA and 
VAC$_{\rm S}$ solutions studied in \cite{BKS4}. 

For an electron energy threshold of 5 MeV the best fit value of the first
moment of the CC electron spectrum $\langle T\rangle$ in the case of the RSFP 
is systematically up-shifted compared to the SSM prediction by about (55
-- 75) keV. The values of the shift $\langle\Delta T\rangle$ allowed at 
90\% c.l. range between $-110$ and +230 keV, which is about three times 
the expected combined calculational and measurement uncertainty, $\pm 96$
keV \cite{BKS4}. For the electron threshold energy of 8 MeV the predicted 
range of $\langle \Delta T\rangle$ is somewhat larger, between $-180$ 
and +215 keV. Therefore in principle SNO holds a potential of measuring 
the shift in $\langle T\rangle$ due to non-standard neutrino properties. 
The shifts in $\langle T\rangle$ in the case of the RSFP have partial 
overlap with those predicted in the case of neutrino oscillations in 
vacuum or in matter \cite{BKS4}. Our predicted values of the shift in the
second moment $\sigma$ are smaller than those in $\langle T\rangle$. 
Their values allowed at 90\% c.l. range between $-70$ and $82$ keV, 
which has to be compared with the $1\sigma$ uncertainty of $\pm 44$ keV. 
Again, there is a partial overlap between our predictions for $\Delta 
\sigma$ and those in the case of neutrino oscillations \cite{BKS4}, but 
there are also rather large regions of no overlap, especially in the case 
of VAC$_{\rm S}$ and VAC$_{\rm L}$ oscillation solutions.  

We therefore conclude that the possibility of experimentally disentangling  
the two types of solutions to the solar neutrino 
problem (neutrino oscillations and magnetic moments) depends to a large 
extent on where in the allowed region the neutrino parameters lie. 
Given the uncertainties in the calculations and the expected uncertainties
in the experimental results, the unambiguous discrimination 
on the basis of the average rates and electron spectrum distortions 
appears to be difficult. 

On the other hand, the RSFP mechanism can lead to time dependence of the 
solar neutrino signal due to the variability of the solar magnetic field 
strength. Such time dependence should be manifest in the CC reaction rate  
and electron spectrum, but should not be observable in the NC rate provided 
that neutrinos are Majorana particles \cite{Akh2,BG1}. In particular, 
the CC signal can have an 11-year periodicity related to the solar activity 
cycle \cite{rev}. There may also be seasonal variations of the CC signal due 
to the surface equatorial gap in the toroidal magnetic field of the sun
and a non-zero angle between the solar equatorial plane and the earth's 
ecliptic \cite{VVO}. These seasonal variations are expected to be different 
from those expected in the case of VAC$_{\rm S}$ and VAC$_{\rm L}$ oscillation 
solutions. However, 
the gap in the magnetic field seen at the surface of the sun may not be
present in the inner regions where the RSFP effectively takes the place.
It is therefore difficult to make an unambiguous prediction on whether or
not the seasonal variations of the CC signal should take place in the case of 
the RSFP scenario. 

If  both transition magnetic moments and mixing of massive neutrinos 
are present, the combined action of the RSFP and neutrino oscillations may 
lead to an observable flux of solar $\bar{\nu}_e$'s \cite{LM,Akh3} provided 
that the mixing angle is not too small, $\sin 2\theta_0\aprge 0.1$. 
SNO can detect $\bar{\nu}_e$ through the CC reaction $\bar{\nu}_e+d\to  
n+n+e^+$ (for a recent discussion of the $\bar{\nu}_e$ signal in SNO see
the second reference in \cite{ALP1}).

Thus, the possible time dependence of the CC signal and spectrum distortion 
together with the time independence of the NC signal in the case of the
magnetic moment solution remains the best hope for a discrimination 
between the neutrino magnetic moment and oscillation scenarios. For a
hybrid solution (neutrino magnetic moment plus flavour mixing) the smoking 
gun signature would be an observation of $\bar{\nu}_e$'s from the sun.
 
\vspace{0.2cm}  
\noindent
We are grateful to K. Kubodera for useful correspondence.  
The work of E. A. was supported by Funda\c{c}\~ao para a Ci\^encia e a 
Tecnologia through the grant PRAXIS XXI/BCC/16414/98 and also 
in part by the TMR network grant ERBFMRX-CT960090 of the European Union.

\end{document}